\documentclass{article}

\PassOptionsToPackage{numbers, compress}{natbib}
\usepackage[preprint]{neurips_2022}

\usepackage[utf8]{inputenc} % allow utf-8 input
\usepackage[T1]{fontenc}    % use 8-bit T1 fonts
\usepackage{hyperref}       % hyperlinks
\usepackage{url}            % simple URL typesetting
\usepackage{booktabs}       % professional-quality tables
\usepackage{amsfonts}       % blackboard math symbols
\usepackage{nicefrac}       % compact symbols for 1/2, etc.
\usepackage{microtype}      % microtypography
\usepackage{xcolor}         % colors
\usepackage[pdftex]{graphicx}
\usepackage{subfig}
\usepackage{caption}
\usepackage{varwidth}
\usepackage{enumitem}
\usepackage{amsmath}
\usepackage{wrapfig}
\usepackage{multicol}
\usepackage{dsfont}
\newsavebox\tmpbox

\newcommand{\perdoor}{\texttt{PerDoor}}

\title{\perdoor: Persistent Non-Uniform Backdoors in Federated Learning using Adversarial Perturbations}

\author{%
Manaar Alam\\
Center for Cyber Security\\
New York University Abu Dhabi\\
Abu Dhabi, UAE\\
\texttt{alam.manaar@nyu.edu} \\
\And
Esha Sarkar \\
Tandon School of Engineering\\
New York University\\
New York, USA\\
\texttt{esha.sarkar@nyu.edu} \\
\And
Michail Maniatakos\\
Center for Cyber Security\\
New York University Abu Dhabi\\
Abu Dhabi, UAE\\
\texttt{michail.maniatakos@nyu.edu} \\
}

\begin{document}

\maketitle

\begin{abstract}
Federated Learning (FL) enables numerous participants to train deep learning models collaboratively without exposing their personal, potentially sensitive data, making it a promising solution for data privacy in collaborative training. The distributed nature of FL and unvetted data, however, makes it inherently vulnerable to \textit{backdoor attacks}: In this scenario, an adversary injects backdoor functionality into the centralized model during training, which can be triggered to cause the desired misclassification for a specific adversary-chosen input. A range of prior work establishes successful backdoor injection in an FL system; however, \textit{these backdoors are not demonstrated to be long-lasting}. The backdoor functionality does not remain in the system if the adversary is removed from the training process since the centralized model parameters continuously mutate during successive FL training rounds. %The backdoor functionality in a production FL system may not even persist until deployment. 
Therefore, in this work, we propose \perdoor, a persistent-by-construction backdoor injection technique for FL, driven by \textit{adversarial perturbation} and targeting parameters of the centralized model that \textit{deviate less in successive FL rounds} and \textit{contribute the least to the main task accuracy}. An exhaustive evaluation considering an image classification scenario portrays on average $10.5\times$ persistence over multiple FL rounds compared to traditional backdoor attacks. Through experiments, we further exhibit the potency of \perdoor{} in the presence of state-of-the-art backdoor prevention techniques in an FL system. Additionally, the operation of adversarial perturbation also assists \perdoor{} in developing \textit{non-uniform trigger patterns} for backdoor inputs compared to \textit{uniform triggers} (with fixed patterns and locations) of existing backdoor techniques, which are prone to be easily mitigated.
\end{abstract}

\section{Introduction}\label{sec:intro}
Federated Learning (FL)~\cite{DBLP:journals/corr/KonecnyMYRSB16,DBLP:conf/aistats/McMahanMRHA17}, a decentralized deep learning framework, enables massively distributed collaborative training with thousands or even millions of participants~\cite{DBLP:journals/corr/abs-1902-01046,DBLP:journals/corr/abs-1811-03604}. FL is an iterative protocol where a central server broadcasts the current global model to a random subset of participants. The participants then update the received model locally using their own data and transmit the updated model back to the central server, which aggregates the updates into a new global model. FL provides a promising framework where participants need not share private information with the service provider or other participants, promising for settings where data privacy is desired.

Most FL settings assume that the central server can not verify the training data of participating clients and can not control their training process. Hence, despite its benefits, FL is shown to be vulnerable to so-called \textit{backdoor attacks} where an adversary can manipulate local models of a subset of participants and inject backdoor functionalities into the aggregated global model, producing target incorrect predictions for inputs chosen by the adversary.~\cite{DBLP:conf/aistats/BagdasaryanVHES20,DBLP:journals/corr/abs-1911-07963,DBLP:conf/nips/WangSRVASLP20,DBLP:conf/iclr/XieHCL20}. Though the existing attacks deliver very high backdoor accuracy immediately after the iteration when the backdoor is injected into the global model, they are not designed to maintain the persistence of the backdoor functionality, which largely depends on the poisoned parameters\footnote{Parameters of neural network consist of `weights' and `biases'. In this paper, we treat these two identically.} of the global model. The central server aggregates global model using local updates of other benign participants in successive training iterations, resulting in continuous mutation of global model parameters. As a result, if the adversary is removed from training iterations, the global model will gradually unlearn backdoor task, and the backdoor functionality will diminish as FL training continues. In this work, we uncover a new vulnerability that ensures an adversarial impact on FL setting, even after an active adversary is removed. We propose \perdoor{}, a \textit{persistent-by-construction} backdoor injection methodology for FL, which aims to address the challenge of backdoor persistence in existing works. \perdoor{} targets parameters of the global model that deviate less in successive FL rounds to implant the backdoor functionality for better persistence. In addition, \perdoor{} does not tamper with parameters that contribute most to the main task accuracy of the global model and has no distinguishable impact to other benign model updates.

\begin{figure}[!t]
    \centering
    \subfloat[\label{fig:overview_traditional}]{\includegraphics[width=0.5\linewidth]{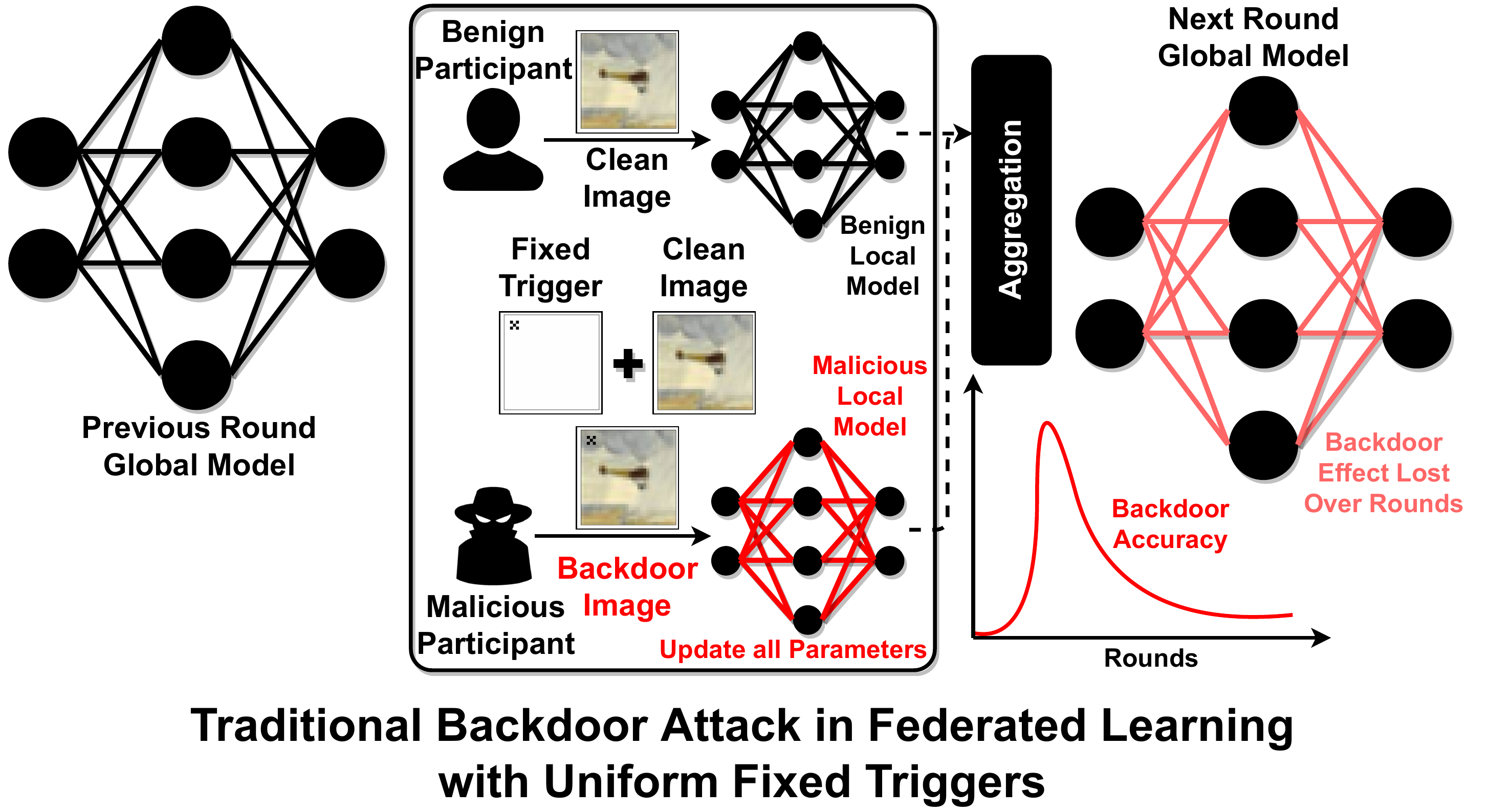}}
    \subfloat[\label{fig:overview_perdoor}]{\includegraphics[width=0.5\linewidth]{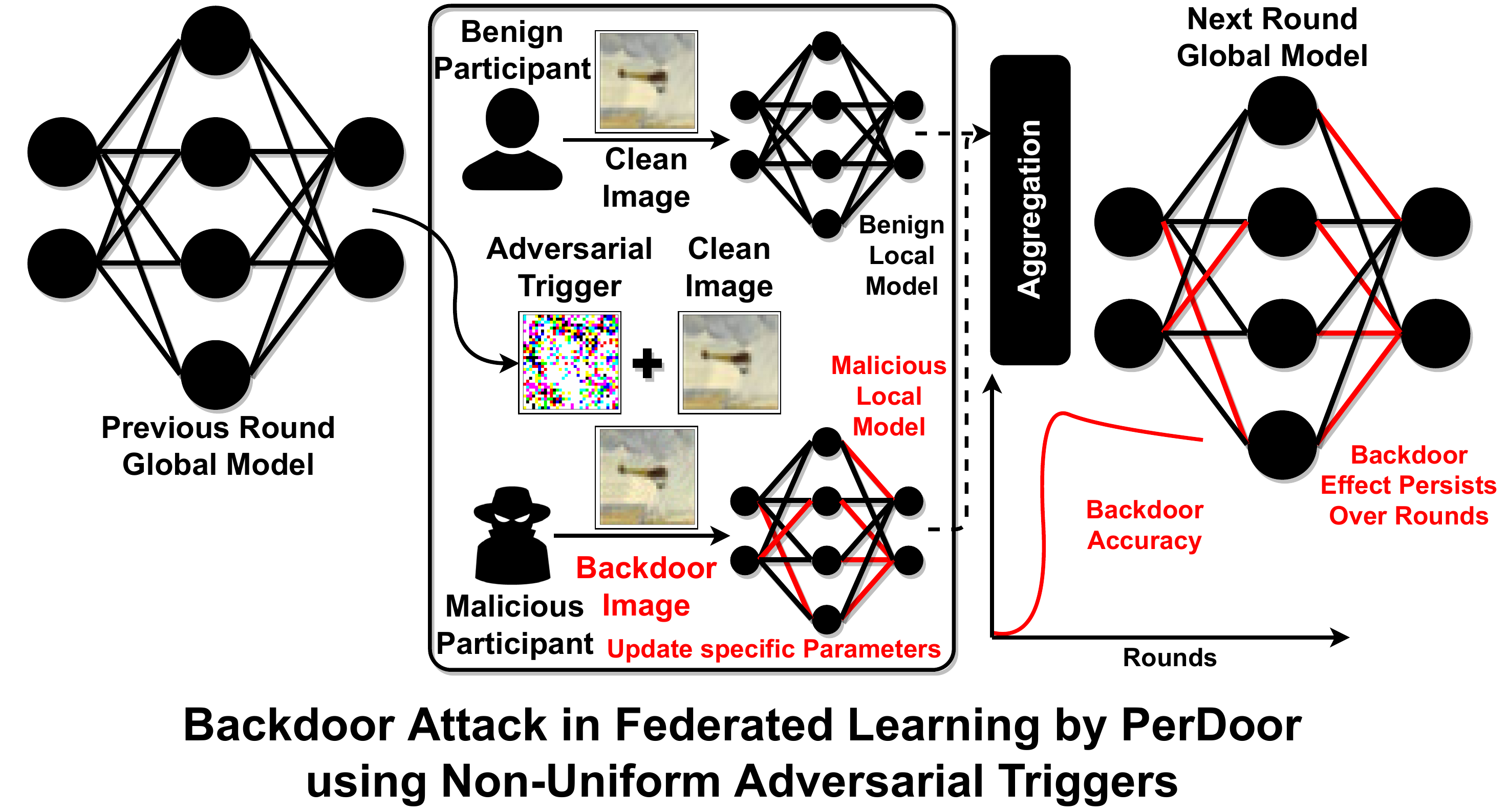}}
    \caption{(a) Traditional backdoor attacks in FL with \textit{fixed uniform triggers} inject backdoors into the global model by modifying \textit{all} its parameters, which gradually unlearns the backdoor functionality over FL rounds once the adversary is removed from training process. (b) \perdoor{} with \textit{adversarially generated triggers} from global model injects backdoors into it by modifying \textit{specific} parameters, making it more persistent over FL rounds.}
    \label{fig:perdoor}
\end{figure}

\perdoor{} uses the principle of \textit{adversarial attacks} to generate backdoor triggers\footnote{A secret pattern that helps misclassify altered inputs to an incorrect, targeted label.} because of their ability to misclassify human-imperceptible perturbed examples to the desired class with high efficiency~\cite{DBLP:journals/corr/abs-1810-00069}. Adversarial attacks have previously been used to generate efficient backdoor triggers for traditional deep learning systems~\cite{DBLP:conf/issta/ZhangDTGYJ21}; however, the efficacy of such a method in a massively distributed FL framework has not been explored previously. The application of adversarial attacks also aids \perdoor{} in generating non-uniform trigger patterns for different inputs compared to the fixed uniform trigger patterns of existing backdoor attacks in FL~\cite{DBLP:conf/aistats/BagdasaryanVHES20,DBLP:conf/aaai/OzdayiKG21,DBLP:conf/iclr/XieHCL20}, which are prone to be easily mitigated~\cite{DBLP:journals/corr/abs-2003-03675}. A brief illustration of \perdoor{} is shown in Figure~\ref{fig:overview_perdoor} compared to the traditional backdoor attacks in FL shown in Figure~\ref{fig:overview_traditional}.

Existing defenses against backdoor attacks in FL primarily aim to prevent it instead of detecting it, as it has been shown that detecting backdoors is challenging without breaking the privacy of participants~\cite{DBLP:conf/aistats/BagdasaryanVHES20,DBLP:conf/sp/MelisSCS19,DBLP:journals/corr/abs-1812-00910}. Moreover,  detecting backdoors in a model is proven to be an \textit{NP-Hard} problem~\cite{DBLP:conf/nips/WangSRVASLP20}. We consider two state-of-the-art backdoor prevention methods in FL, Krum~\citep{DBLP:conf/nips/BlanchardMGS17}  and FoolsGold~\citep{DBLP:conf/raid/FungYB20}, which use byzantine-tolerant aggregation methods and evaluate the effectiveness of \perdoor{} in presence of such defenses. Experimental evaluation considering an image classification scenario demonstrates that \perdoor{} can achieve, on average, $10.5\times$ persistence over successive iterations of FL training compared to traditional backdoor injection methods when the backdoor is injected at the same round. We also show that \perdoor{} can achieve high persistence even in the presence of state-of-the-art defenses.

\noindent \textbf{Our Contribution:}
We summarize our main contributions as follows: \textbf{(1)} We propose \perdoor{}, a backdoor injection method in FL, which is persistent-by-construction. \perdoor{} targets parameters of the centralized global model that deviate less in successive FL rounds and contribute the least to the main task accuracy. Targeting such parameters for injecting backdoors instead of affecting all parameters makes \perdoor{} more persistent than traditional backdoor injection methods in FL. \textbf{(2)} We use the principle of adversarial attacks to generate backdoor triggers for \perdoor{}, which to the best of our knowledge, has not been explored in an FL framework. The non-uniform triggers generated from adversarial perturbation provide high backdoor accuracy and make it challenging to patch compared to fixed uniform triggers of traditional backdoor injection methods.

\section{Problem Formulation and Objectives}\label{sec:overview}
\noindent \textbf{Brief Overview of Federated Learning:}
Let us consider an FL framework having $n$ participants $\{p_1, \dots, p_n\}$ and a central aggregation server that collaboratively train a global model $\mathcal{G}$. In a training round $r \in \{1, \dots, \mathcal{R}\}$, the central server randomly selects a (typically small) subset of $m$ participants $\mathcal{S}_m$ and broadcasts  the current global model $\mathcal{G}^{r}$. Starting from $\mathcal{G}^{r}$, each selected participant $p_i \in \mathcal{S}_m$ trains respective local models $\mathcal{L}^{r+1}_{p_i}$ using its private data $\mathcal{D}_{p_i}$ by running a standard optimization algorithm, and transmits the difference $\mathcal{L}^{r+1}_{p_i} - \mathcal{G}^r$ back to the central server, which aggregates the received updates into the global model $\mathcal{G}^{r+1}$. In this work, without loss of generality, we assume that the central server employs \textit{Federated Averaging} (FedAvg)~\cite{DBLP:conf/aistats/McMahanMRHA17} aggregation method as it is commonly applied in FL~\cite{DBLP:conf/ccs/BonawitzIKMMPRS17,DBLP:conf/sp/FereidooniMMMMN21,DBLP:conf/aistats/McMahanMRHA17,DBLP:conf/nips/SmithCST17} and related work on backdoor attacks in FL~\cite{DBLP:conf/aistats/BagdasaryanVHES20,DBLP:conf/raid/FungYB20,DBLP:journals/corr/abs-2101-02281,DBLP:conf/nips/WangSRVASLP20}\footnote{We also consider Krum and FoolsGold for attack evaluation in Section~\ref{sec:results}}. Under FedAvg, the global model is updated by performing a weighted average on the received updates as:
\begin{equation}
    \mathcal{G}^{r+1} = \mathcal{G}^{r} + \sum_{p_i \in \mathcal{S}_m}\frac{n_{p_i}}{n_{\mathcal{S}_m}}(\mathcal{L}^{r+1}_{p_i} - \mathcal{G}^r)
\end{equation}
where $n_{p_i} = |\mathcal{D}_{p_i}|$ and $n_{\mathcal{S}_m} = \sum_{p_i \in \mathcal{S}_m} n_{p_i}$ is the total number of training data used at the selected round. Learning does not stop even after the model converges. Participants, throughout their deployment, continuously update the global model. A malicious participant thus always has an opportunity to be selected and influence the global model.

\noindent \textbf{Threat Model:}
Following recent studies~\cite{DBLP:journals/corr/abs-1911-07963,DBLP:conf/nips/WangSRVASLP20}, we consider a scenario where a fixed pool of malicious participants can only conduct the attack in certain FL rounds when the central server randomly selects them. This participating pattern of adversaries is also known as the \textit{random sampling} strategy\footnote{The authors in~\cite{DBLP:journals/corr/abs-1911-07963,DBLP:conf/nips/WangSRVASLP20} also proposed a \textit{fixed-frequency} strategy where a participant is selected at every $f$ round, $f$ being a fixed value. The \textit{random sampling} strategy is considered to be more practical in a realistic robust FL framework. However, the proposed approach is independent of the participating strategy of the attackers.}. Multiple adversaries may participate in a single FL round in such a scenario. While we only consider an independent adversary in this work, collusion can further strengthen the attack as demonstrated in~\cite{DBLP:conf/iclr/XieHCL20}.

\noindent \textit{Capability of the Adversary:}
Following existing work on backdoor attacks in FL~\cite{DBLP:conf/aistats/BagdasaryanVHES20,DBLP:conf/nips/BlanchardMGS17,DBLP:journals/corr/abs-2101-02281}, we consider that an adversary $\mathcal{A}$ has full control over the malicious participants. However, $\mathcal{A}$ has no control over other benign participants nor has access to their data or local updates. We assume that $\mathcal{A}$ has complete knowledge of the central aggregation server and deployed defense, if any. More precisely, $\mathcal{A}$ knows the algorithms used in the central server along with all configuration parameters but can not tamper with them.

\noindent \textit{Goal of the Adversary:} The objectives of $\mathcal{A}$ are four-fold: \textbf{(1)} \textit{Impact:} $\mathcal{A}$ utilizes the broadcasted global model at round $r$, i.e., $\mathcal{G}^r$, to construct a malicious local model $\hat{\mathcal{L}}^{r+1}_{\mathcal{A}}$ using a set of \textit{trigger inputs} $\mathcal{T}$ targeting an incorrect chosen label. $\mathcal{A}$ aims to obtain desired incorrect outputs for the set $\mathcal{T}$ from $\hat{\mathcal{G}}^{r+1}$ at the subsequent round, even after the central server aggregates $\hat{\mathcal{L}}^{r+1}_{\mathcal{A}}$ with local updates of other benign participants. \textbf{(2)} \textit{Stealthiness:} $\mathcal{A}$ aims to construct $\hat{\mathcal{L}}^{r+1}_{\mathcal{A}}$ as indistinguishable as possible from the local models of other benign participants to make it difficult for the central server to identify any malicious activity. $\mathcal{A}$ aims to keep the similarity of $\hat{\mathcal{L}}^{r+1}_{\mathcal{A}}$ with other benign models greater than a threshold that can be estimated by comparing $\hat{\mathcal{L}}^{r+1}_{\mathcal{A}}$ to $\mathcal{G}^r$ as $\mathcal{A}$ does not have access to the local updates of other benign participants. \textbf{(3)} \textit{Persistence:} $\mathcal{A}$ aims to create $\mathcal{T}$ and construct $\hat{\mathcal{L}}^{r+1}_{\mathcal{A}}$ so that $\mathcal{A}$ can obtain desired incorrect outputs for the set $\mathcal{T}$ from $\hat{\mathcal{G}}^{r+1}$ over multiple subsequent rounds, even after numerous aggregations with local updates of other benign participants. \textbf{(4)} \textit{Non-Uniformity:} $\mathcal{A}$ aims to create the trigger inputs in $\mathcal{T}$ so that the trigger patterns and locations are different for different inputs compared to the fixed uniform nature of existing triggers, making it difficult for the central server to patch global models based on the knowledge of trigger patterns.

\noindent \textit{Poisoning Method:} Following existing works on backdoor attacks in FL~\cite{DBLP:conf/aistats/BagdasaryanVHES20,DBLP:journals/corr/abs-1911-07963,DBLP:conf/nips/WangSRVASLP20}, we consider that $\mathcal{A}$ carries out \textit{single-shot model replacement}\footnote{We also consider \textit{continuous poisoning} method to evaluate \perdoor{} against traditional backdoor injection methods in FL later in Section~\ref{sec:results}} attack to inject backdoors in $\hat{\mathcal{G}}^{r+1}$ by transmitting back $\frac{n_{\mathcal{S}_m}}{n_{\mathcal{A}}}(\hat{\mathcal{L}}^{r+1}_{\mathcal{A}} - \mathcal{G}^r) + \mathcal{G}^r$ instead of $\mathcal{G}^r$ to the central server.

\section{Crafting Persistent and Non-Uniform Backdoors using \perdoor{}}\label{sec:perdoor}
The principle idea of \perdoor{} is to target parameters of the centralized global model that deviate less in successive FL rounds and contribute the least to main task accuracy for injecting backdoors. Let the parameter set of the global model $\mathcal{G}^r$ at a particular round $r$ be denoted as $\theta^{r}$. To analyze the deviation of each parameter in $\theta^{r}$ over multiple rounds, an adversary $\mathcal{A}$ monitors $\theta^{r}$ whenever the central server selects it during training and broadcasts $\mathcal{G}^r$. Let us consider a set $\mathcal{W}$ consisting of the rounds when $\mathcal{A}$ is selected during training before injecting backdoors. $\mathcal{W}$ is termed as the analysis window for $\mathcal{A}$ to investigate the target parameters for persistent backdoor injection. Let us denote $\theta_\Delta$ be the set of parameters that deviate less over successive FL rounds, which can be computed as:
\begin{equation}
    \theta_\Delta = \mathds{1}_{\widehat{var}(\{\theta^{r}\}_{r \in \mathcal{W}}) < \mathbf{t_{\Delta}}}
\end{equation}
where $\widehat{var}(\cdot)$ is the estimated variance\footnote{Note that the variance is computed inter rounds for each parameter. It is not an intra-round variance.} and $\mathbf{t_{\Delta}}$ works as a hyperparameter, controlling both stealthiness through the impact on main task accuracy and efficiency through the impact on backdoor accuracy. A higher value of $\mathbf{t_{\Delta}}$ can impact more parameters and thus has a higher chance of affecting main task accuracy. On the contrary, a lower value of $\mathbf{t_{\Delta}}$ impacts fewer parameters but may not inject backdoors efficiently with a high backdoor accuracy.

Let us define $\mathcal{I}_{l}^{r}$ be the contribution of each parameters in the $l$-th layer of $\mathcal{G}^r$ on main task accuracy. Let us also consider $\mathcal{X} = \{(x_1, y_1), (x_2, y_2), \dots, (x_d,y_d)\}$ as the set of benign inputs to which $\mathcal{A}$ has access. The $\mathcal{I}_{l}^{r}$ can be computed as:
\begin{equation}
    \mathcal{I}_{l}^{r} = \frac{1}{|\mathcal{X}|}\sum_{(x, y) \in \mathcal{X}}\nabla_{\theta_{l}^{r}}[\mathcal{G}^r(x)]_{y}
\end{equation}
where $\nabla_{\theta_{l}^{r}}[\mathcal{G}^r(x)]_{y}$ denote the gradient of $\mathcal{G}^r$ on input $x$ considering the label $y$ with respect to the parameters in the $l$-th layer of $\mathcal{G}^r$. Let us denote $\theta_{\sigma_l}^{r}$ be the set of parameters on the $l$-th layer and $\theta_{\sigma}^{r}$ be the set of all parameters of $\mathcal{G}^r$, respectively, that contribute the least to the main task accuracy, which can be computed as:

\begin{minipage}{0.5\linewidth}
\begin{equation}
    \theta_{\sigma_l}^{r} = \mathds{1}_{\mathcal{I}_{l}^{r} < \widehat{mean}(\mathcal{I}_{l}^{r})}
\end{equation}
\end{minipage}%
\begin{minipage}{0.5\linewidth}
\begin{equation}
    \theta_{\sigma}^r = \bigcup_{l}\theta_{\sigma_l}^{r}
\end{equation}
\end{minipage}
where $\widehat{mean}(\cdot)$ is the estimated mean. The backdoor parameters $\theta_\mathcal{B}$ that $\mathcal{A}$ intends to target for injecting backdoor on $\mathcal{G}^{r+1}$ can be computed as: $\theta_\mathcal{B} = \theta_\Delta \cap \theta_{\sigma}^r$.

\perdoor{} uses \textit{adversarial perturbations} on $\mathcal{G}^r$ to generate backdoor triggers because of its ability to misclassify perturbed examples to the desired class with high efficiency~\cite{DBLP:journals/corr/abs-1810-00069}. The adversarial perturbation also generates non-uniform backdoor triggers for different inputs. A brief overview of adversarial example generation is provided in Appendix~\ref{sec:adv_attack}. However, the adversarial examples alone fail to provide desired persistence over successive FL rounds as the efficacy of such attacks depends on the decision boundary (i.e., parameters) of the target model, which continuously mutates over consecutive FL rounds. Hence, $\mathcal{A}$ adjusts the backdoor parameters $\theta_\mathcal{B}$ of the local model to construct $\hat{\mathcal{L}}^{r+1}_{\mathcal{A}}$ for implanting the effect of adversarial examples for better persistence. Let us consider the set of adversarial backdoor images be $\hat{\mathcal{X}} = \{(\hat{x}_1, \hat{y}), (\hat{x}_2, \hat{y}), \dots, (\hat{x}_f, \hat{y})\}$, generated using a small subset of benign examples $\{(x_1, y_1), (x_2, y_2), \dots, (x_f, y_f)\}$ in $\mathcal{X}$ for the target label $\hat{y}$. The goal of $\mathcal{A}$ is to construct $\hat{\mathcal{L}}^{r+1}_{\mathcal{A}}$ by only adjusting $\theta_\mathcal{B}$ so that its loss is minimized for the examples in $\hat{\mathcal{X}}$. The parameter modification is performed iteratively over $\mathcal{K}$ local iterations as:
\begin{equation}\label{eq:perdoor}
    [\hat{\mathcal{L}}^{r+1}_{\mathcal{A}}]_{\theta_{\mathcal{B}}}^{k+1} = [\hat{\mathcal{L}}^{r+1}_{\mathcal{A}}]_{\theta_{\mathcal{B}}}^{k} - \text{clip}_{\delta}\left(\delta \cdot \frac{1}{|\hat{\mathcal{X}}|}\sum_{(x, y)\in\hat{\mathcal{X}}}\nabla_{\theta_{\mathcal{B}}}\mathcal{J}(\hat{\mathcal{L}}^{r+1}_{\mathcal{A}}, x, y)\right)
\end{equation}
where $[\hat{\mathcal{L}}^{r+1}_{\mathcal{A}}]_{\theta_{\mathcal{B}}}^{k}$ denotes the values of backdoor parameters $\theta_{\mathcal{B}}$ at $k$-th local iteration, initial value of $\hat{\mathcal{L}}^{r+1}_{\mathcal{A}}$ is $\mathcal{G}^r$, $\mathcal{J}(\hat{\mathcal{L}}^{r+1}_{\mathcal{A}}, x, y)$ denotes the loss for $\hat{\mathcal{L}}^{r+1}_{\mathcal{A}}$ considering input $x$ and corresponding label $y$, $\nabla_{\theta_{\mathcal{B}}}$ denotes the gradient with respect to the backdoor parameters $\theta_{\mathcal{B}}$, and $\delta$ (termed as backdoor strength) works as a parameter controlling both stealthiness through the impact on main task accuracy and efficiency through the impact on backdoor accuracy, $\text{clip}_{\delta}(\cdot)$ denotes the function regulating parameter modification within $\delta$-ball of the original parameters. A higher value of $\delta$ introduces large modifications in $\hat{\mathcal{L}}^{r+1}_{\mathcal{A}}$, which may affect main task accuracy. On the contrary, a lower value of $\delta$ may not inject backdoors efficiently with a high backdoor accuracy. As discussed in Section~\ref{sec:overview}, an optimum value of $\delta$ can be estimated by comparing $\hat{\mathcal{L}}^{r+1}_{\mathcal{A}}$ to $\mathcal{G}^r$.

\section{Experimental Evaluation}\label{sec:results}
\subsection{Experimental Setup}\label{sec:dataset}
\noindent \textbf{Experimental Platform:} We conduct all experiments on a server with a 24-core AMD Ryzen 3960X CPU, NVIDIA GeForce RTX 2080 TI GPU (8GB memory), and 256GB RAM. We used the Python-based Deep Learning framework PyTorch~\cite{DBLP:conf/nips/PaszkeGMLBCKLGA19} for all implementations.

\noindent \textbf{Dataset and Training Configurations:} We consider an image classification task to evaluate the effectiveness of \perdoor{} in FL. Following recent studies on FL~\cite{DBLP:conf/aistats/BagdasaryanVHES20,DBLP:conf/nips/BlanchardMGS17,DBLP:conf/raid/FungYB20,DBLP:conf/nips/WangSRVASLP20}, we use CIFAR-10~\cite{krizhevsky2009learning} as benchmark dataset. We consider an FL framework where the participants collectively train the global model using training data that is independent and identically distributed among the participants. We use VGG-11~\cite{DBLP:journals/corr/SimonyanZ14a} with 9,756,426 parameters as the classifier.

Following the FL training configuration discussed in~\cite{DBLP:conf/aistats/BagdasaryanVHES20}, we train the global model with $n=100$ total participants, 10 of whom are selected randomly in each FL round, i.e., $|\mathcal{S}_m| = 10$. Each chosen participant in a round trains individual local model for 2 epochs with the learning rate of 0.1, as described in~\cite{DBLP:conf/aistats/BagdasaryanVHES20}. The local training is performed using the \textit{stochastic gradient descent} optimization and the \textit{negative log-likelihood loss}. As discussed in Section~\ref{sec:overview}, the central server uses the widely used standard FedAvg aggregation method as baseline implementation. We also consider two state-of-the-art backdoor preventing aggregations in FL, Krum~\citep{DBLP:conf/nips/BlanchardMGS17} and FoolsGold~\citep{DBLP:conf/raid/FungYB20}, and reimplement them for CIFAR-10 and VGG-11 to evaluate the effectiveness of \perdoor{} in presence of such defenses.

\noindent \textbf{Attack Strategy:} We consider $1\%$ of the total participants to be malicious for our evaluation\footnote{An increased number of malicious participants will boost the attack effectiveness, as shown in recent studies on backdoor attacks in FL~\cite{DBLP:conf/aistats/BagdasaryanVHES20,DBLP:conf/raid/FungYB20,DBLP:journals/corr/abs-2101-02281,DBLP:journals/corr/abs-1911-07963,DBLP:conf/nips/WangSRVASLP20,DBLP:conf/iclr/XieHCL20}}. As mentioned in Section\ref{sec:overview}, malicious participants are selected using the \textit{random sampling} strategy following a uniform distribution. We consider two analysis windows for the malicious participants before injecting backdoors: \textbf{(1)} \textit{stable point} $(|\mathcal{W}| = 30)$ - where the global model is close to generalizing the training data, and \textbf{(2)} \textit{volatile point} $(|\mathcal{W}| = 20)$ - where the global model still produces large differences in generalization errors. We considered these two points for evaluating the impact of backdoors when injected at different stages during FL training. The values of $\mathcal{W}$ are empirically selected considering the classifier architecture and the training dataset, which may differ from other architecture and/or datasets. We use the \textit{Basic Iterative Method} (BIM)~\cite{DBLP:conf/iclr/KurakinGB17a} with $L_\infty$-norm to generate adversarial perturbations for backdoor triggers. A brief overview of BIM is provided in Appendix~\ref{sec:adv_attack}. We use the \textit{CleverHans v2.1.0} library~\cite{cleverhans} for implementing BIM. We evaluate the backdoor impact using three different attack strengths $(\epsilon = 0.05, 0.075, 0.1)$ for BIM. We primarily use $\epsilon = 0.1$ for generating the adversarial perturbations unless mentioned otherwise. Without loss of generality, we select two labels randomly as the source (`airplane') and target label (`frog'). The method is equally applicable for other combinations of source and target labels.

\noindent \textbf{Evaluation Metrics:} We consider the following two metrics to evaluate the effectiveness of \perdoor{}: \textbf{(1)} $Acc_{\mathcal{B}}$ (Backdoor Accuracy) indicating the accuracy of a model in the backdoor task, i.e., it is the fraction of trigger images for which the model provides target outputs as chosen by the adversary and \textbf{(2)} $Acc_{\mathcal{M}}$ (Main Task Accuracy) indicating the accuracy of a model in its main task, i.e., it denotes the fraction of benign images for which the model provides correct predictions. The adversary aims to maximize $Acc_{\mathcal{B}}$ and minimize the effect on $Acc_{\mathcal{M}}$.

\subsection{Research Questions}
We construct subsequent experiments to answer the following research questions (RQ) considering the objectives of an adversary as discussed in Section~\ref{sec:overview}:
\begin{itemize}
    \item [\textbf{[RQ1]}] \textit{What is the attack performance of} \perdoor\textit{?} In this RQ, we investigate the effectiveness of \perdoor{} in terms of $Acc_{\mathcal{B}}$ and $Acc_{\mathcal{M}}$. A higher value of $Acc_{\mathcal{B}}$ indicates a successful backdoor injection which should not drop the value of $Acc_{\mathcal{M}}$. We also investigate the stealthiness of such backdoor injection by monitoring similarity of the malicious local model with the broadcasted global model.
    \item [\textbf{[RQ2]}] \textit{How persistent are the backdoors over FL rounds?} In this RQ, we investigate the persistence of \perdoor{} in terms of $Acc_{\mathcal{B}}$ after the adversary stops uploading poisoned updates. A higher value of $Acc_{\mathcal{B}}$ over subsequent FL rounds indicates a persistent backdoor injection.
    \item [\textbf{[RQ3]}] \textit{How does} \perdoor{} \textit{perform against state-of-the-art backdoor defenses?} In this RQ, we investigate the performance of \perdoor{} against two state-of-the-art defense methods in terms of $Acc_{\mathcal{B}}$. A higher value of $Acc_{\mathcal{B}}$ over multiple FL rounds indicates a successful evasion of backdoor defenses.
    \item [\textbf{[RQ4]}] \textit{How diverse are the backdoor trigger patterns for different inputs?} In this RQ, we investigate the non-uniformity of backdoor trigger patterns in \perdoor{} for different input images of the same source label and target label compared to traditional uniform backdoor triggers.
\end{itemize}

\subsection{Experimental Results}
\begin{figure}[!b]
    \centering
    \subfloat[\label{fig:param_no_change}]{\includegraphics[width=0.5\linewidth]{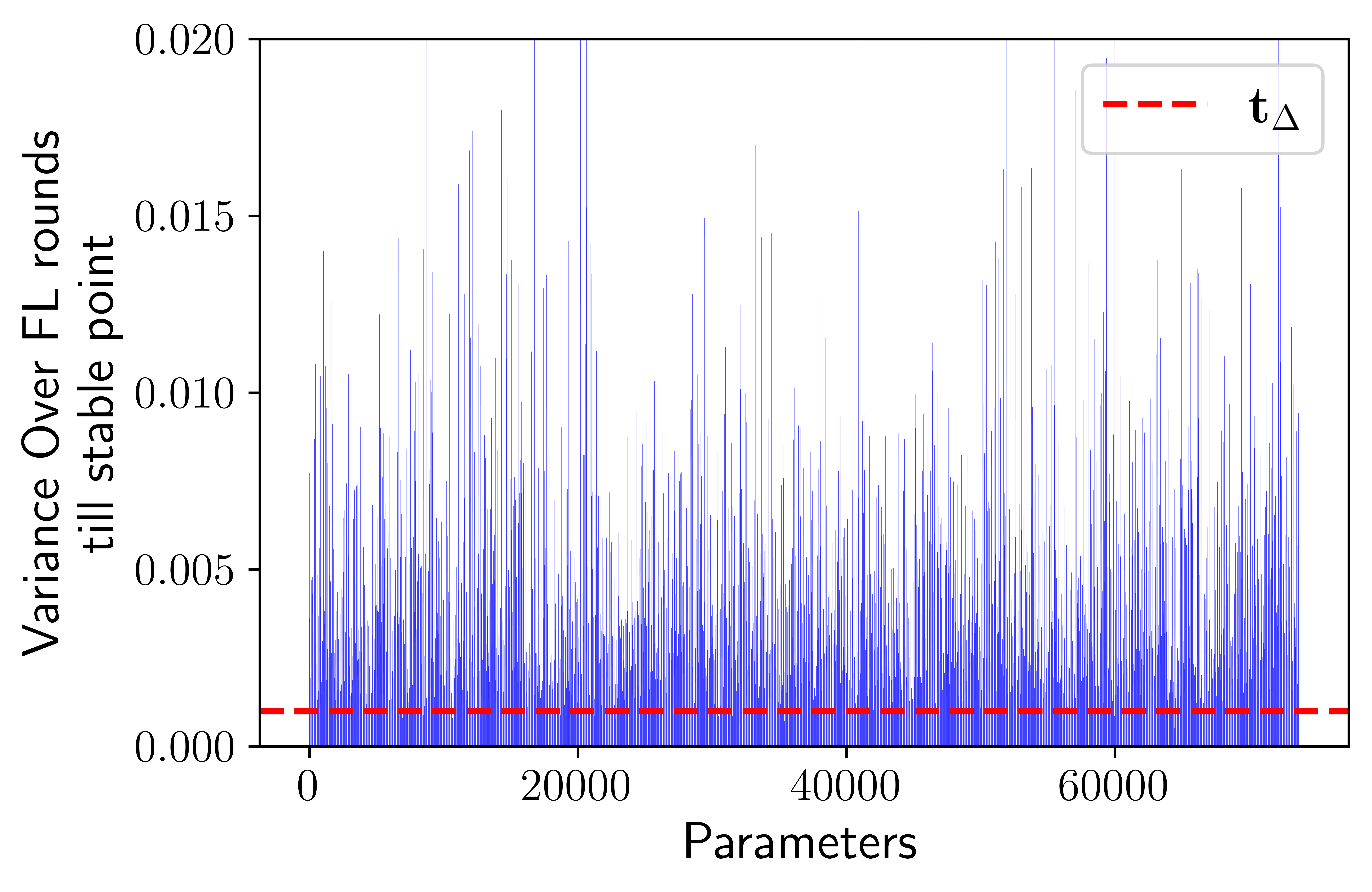}}
    \subfloat[\label{fig:param_important}]{\includegraphics[width=0.5\linewidth, height=4.6cm]{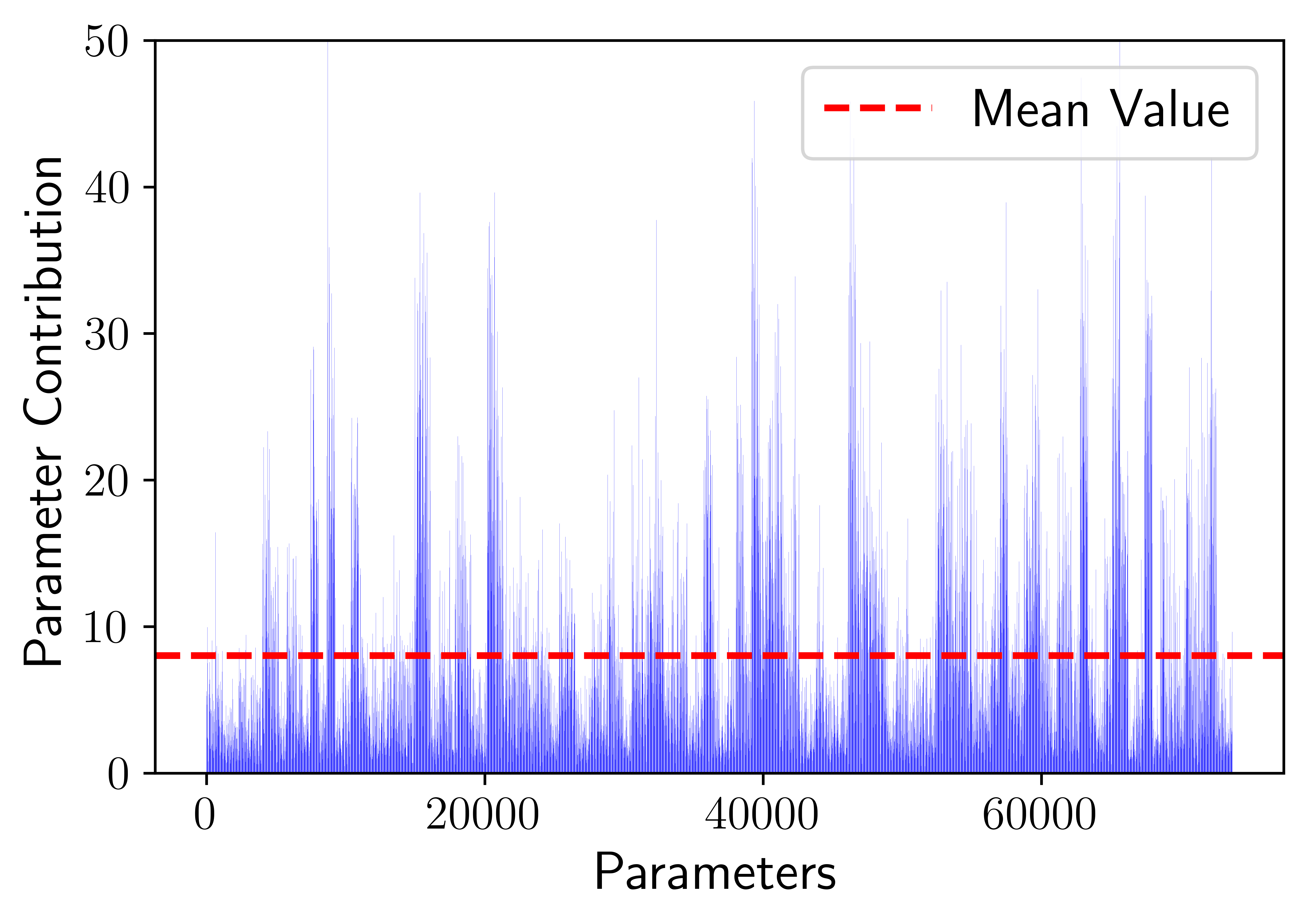}}
    \caption{(a) Variance of parameters in second convolution layer over successive FL rounds till the stable point. (b) Contribution of parameters in second convolution layer towards main task accuracy.}
\end{figure}

We first identify the set of backdoor parameters $\theta_{\mathcal{B}}$ in the global model. Without loss of generality, we assume that the adversary analyzes the broadcasted global model parameters until the stable point of FL rounds for this analysis. For simplicity, we present the variance of each parameter in the second convolution layer of the global model over successive FL rounds till the stable point in Figure~\ref{fig:param_no_change}. In Figure~\ref{fig:param_important}, we present the contribution of each parameter in the second convolution layer with respect to main task accuracy on the global model at the stable point of FL rounds.

The total number of parameters in $\theta_{\mathcal{B}}$ are shown in Table~\ref{table:backdoor_params} for different values of $t_{\Delta}$. We can intuitively observe that as the value of $t_{\Delta}$ decreases, fewer parameters are selected for the backdoor injection. As discussed in Section~\ref{sec:perdoor}, affecting a higher number of parameters poses a possibility of becoming distinguishable from other benign model updates, whereas affecting a lower number of parameters may fail to create the desired backdoor impact. In our further analysis, we have selected $t_{\Delta} = 0.001$, which provides the best performance considering both stealthiness and backdoor impact. The average cumulative variance of $\theta_{\mathcal{B}}$ and the non-backdoor parameters over successive FL rounds after the backdoor injection is shown in Figure~\ref{fig:avg_variance}. Even after the malicious modification through $\theta_{\mathcal{B}}$, we can observe that their average variance is always significantly lower than the average variance of non-backdoor parameters, aiding the persistence of backdoors which we will discuss later.

\begin{table}[!t]
\begin{varwidth}[b]{0.4\linewidth}
\caption{Total number of backdoor parameters for different values of $\mathbf{t_{\Delta}}$.\label{table:backdoor_params}}
\begin{tabular}{cc}
\toprule
$\mathbf{t_{\Delta}}$      & \# backdoor parameters \\ \midrule \midrule
0.01    & 6,263,074     \\ \midrule
0.001   & 4,012,994     \\ \midrule
0.0001  & 186,504       \\ \midrule
0.00001 & 13,580        \\ \bottomrule
\end{tabular}
\end{varwidth}%
\begin{minipage}[t]{0.6\linewidth}
\vskip -2.5cm
\centering
\includegraphics[width=0.85\linewidth]{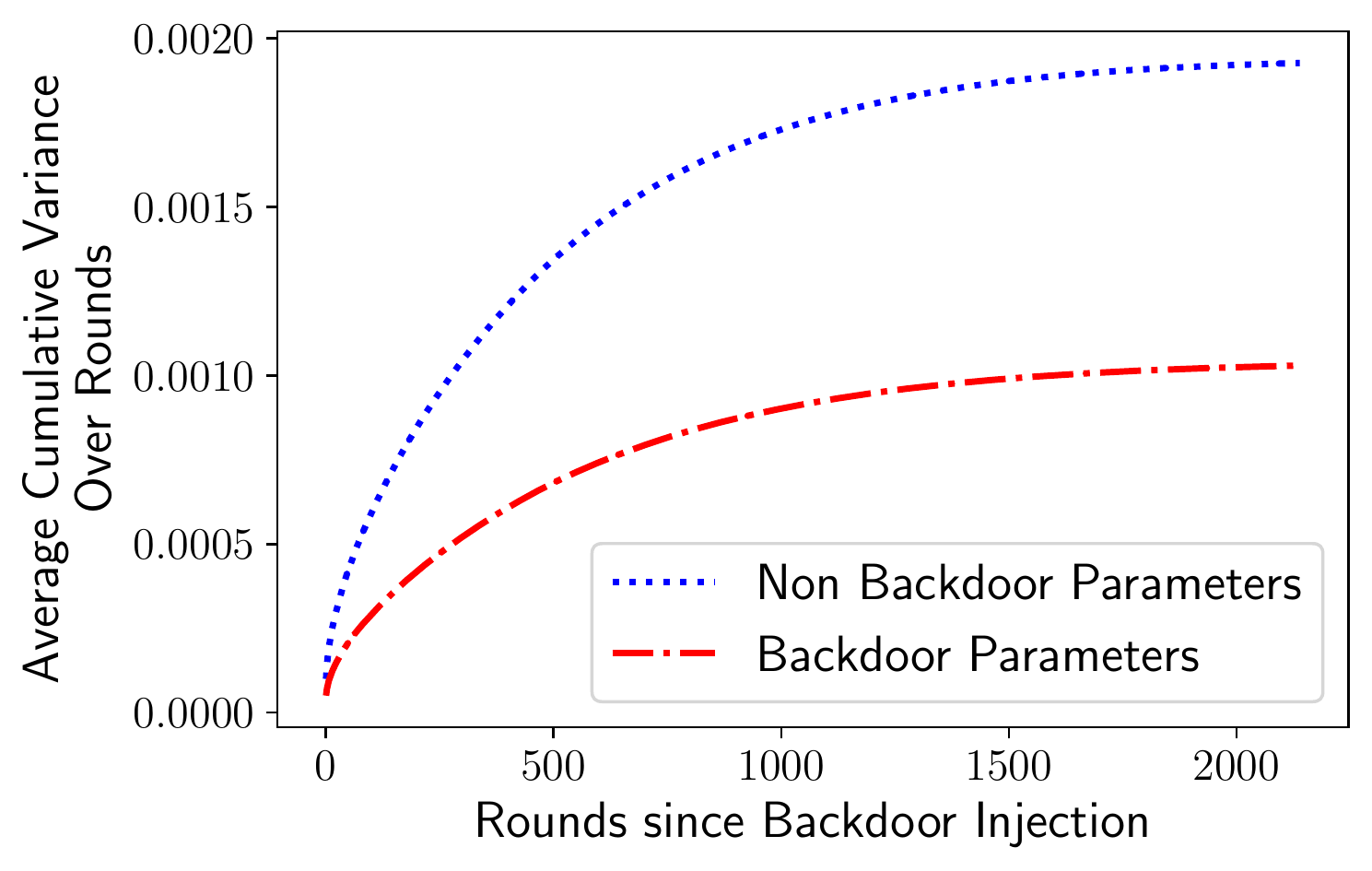}
\captionof{figure}{Average cumulative variance of backdoor and non-backdoor parameters over successive FL rounds.\label{fig:avg_variance}}
\end{minipage}
\end{table}

\noindent \textbf{[RQ1] Attack performance of \perdoor:}
We inject backdoors in the global model with different values of backdoor strength $\delta$ for this analysis. To analyze the similarity of the malicious backdoor model with the broadcasted global model, we use  linear \textit{Central Kernel Alignment} (CKA) analysis~\cite{DBLP:conf/icml/Kornblith0LH19}.
\begin{wrapfigure}{r}{0.5\linewidth}
  \begin{center}
    \includegraphics[width=\linewidth]{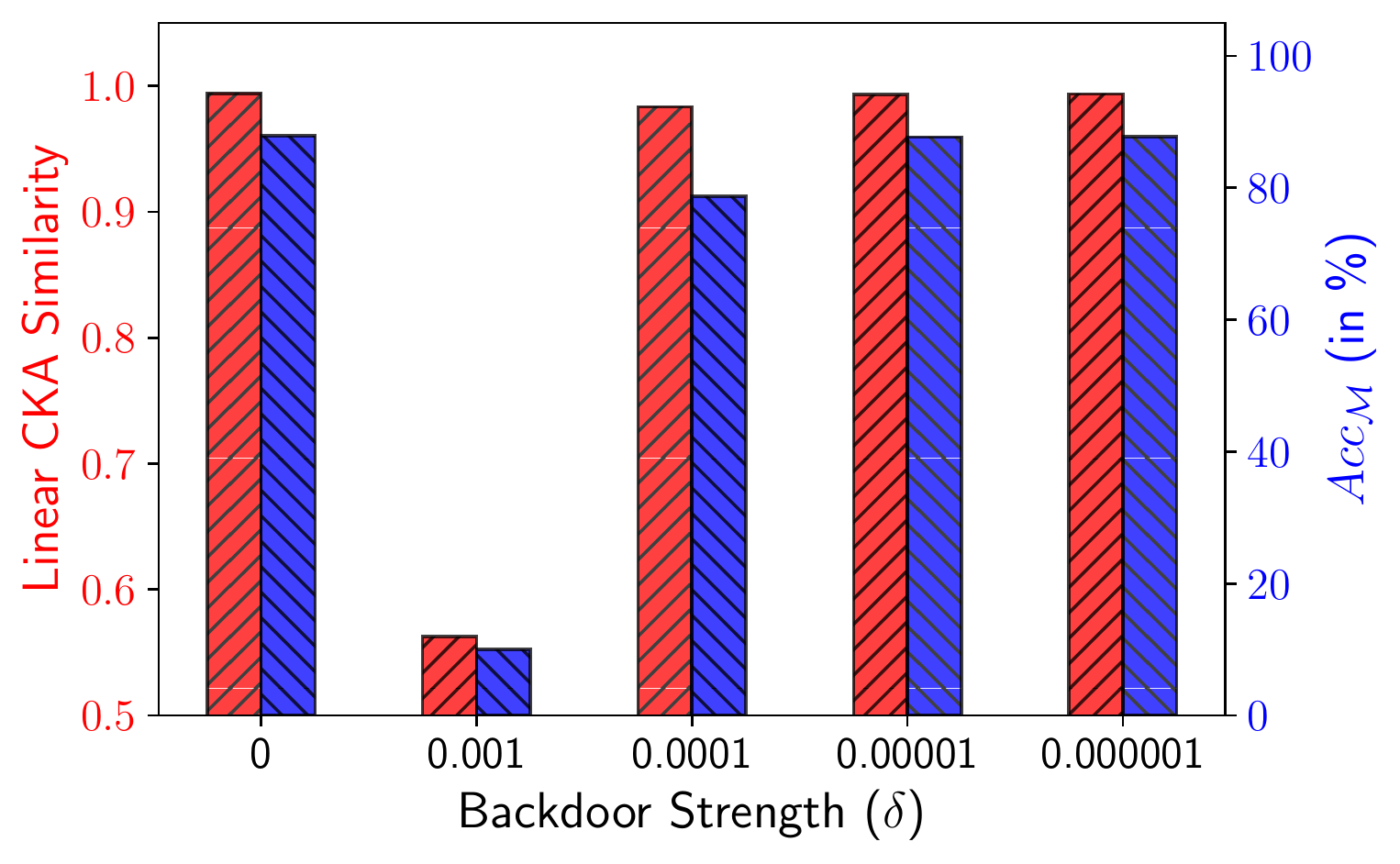}
  \end{center}
  \caption{CKA similarity and main task accuracy $Acc_{\mathcal{M}}$ for different values of backdoor strength $\delta$.\label{fig:cka_acc}}
\end{wrapfigure}
The CKA metric, which lies between $[0, 1]$, measures the similarity between a pair of neural network models, where a higher CKA value indicates substantial similarity. We compute the CKA similarity between $\hat{\mathcal{L}}^{r+1}_{\mathcal{A}}$ and $\mathcal{G}^r$ for different values of $\delta$ and present the result in Figure~\ref{fig:cka_acc}. We can observe that with an increase in the value of $\delta$, the CKA value reduces, indicating a significant drop in the similarity between $\hat{\mathcal{L}}^{r+1}_{\mathcal{A}}$ and $\mathcal{G}^r$. Figure~\ref{fig:cka_acc} also presents $Acc_{\mathcal{M}}$ of the global model $\hat{\mathcal{G}}^{r+1}$ immediately after backdoor injection for corresponding values of $\delta$. We can likewise observe that with an increase in the value of $\delta$, $Acc_{\mathcal{M}}$ also reduces, indicating a substantial influence on the main task accuracy due to backdoor injections. As discussed previously, an appropriate value of $\delta$ can be estimated by comparing the CKA values of $\hat{\mathcal{L}}^{r+1}_{\mathcal{A}}$ to the broadcasted $\mathcal{G}^r$ in successive FL rounds. In our further analysis, we have estimated $\delta = 0.00001$, which provides the best performance considering both stealthiness and backdoor impact.

\begin{figure}[!t]
    \centering
    \subfloat[\label{fig:fedavg_acc}]{\includegraphics[width=0.42\linewidth]{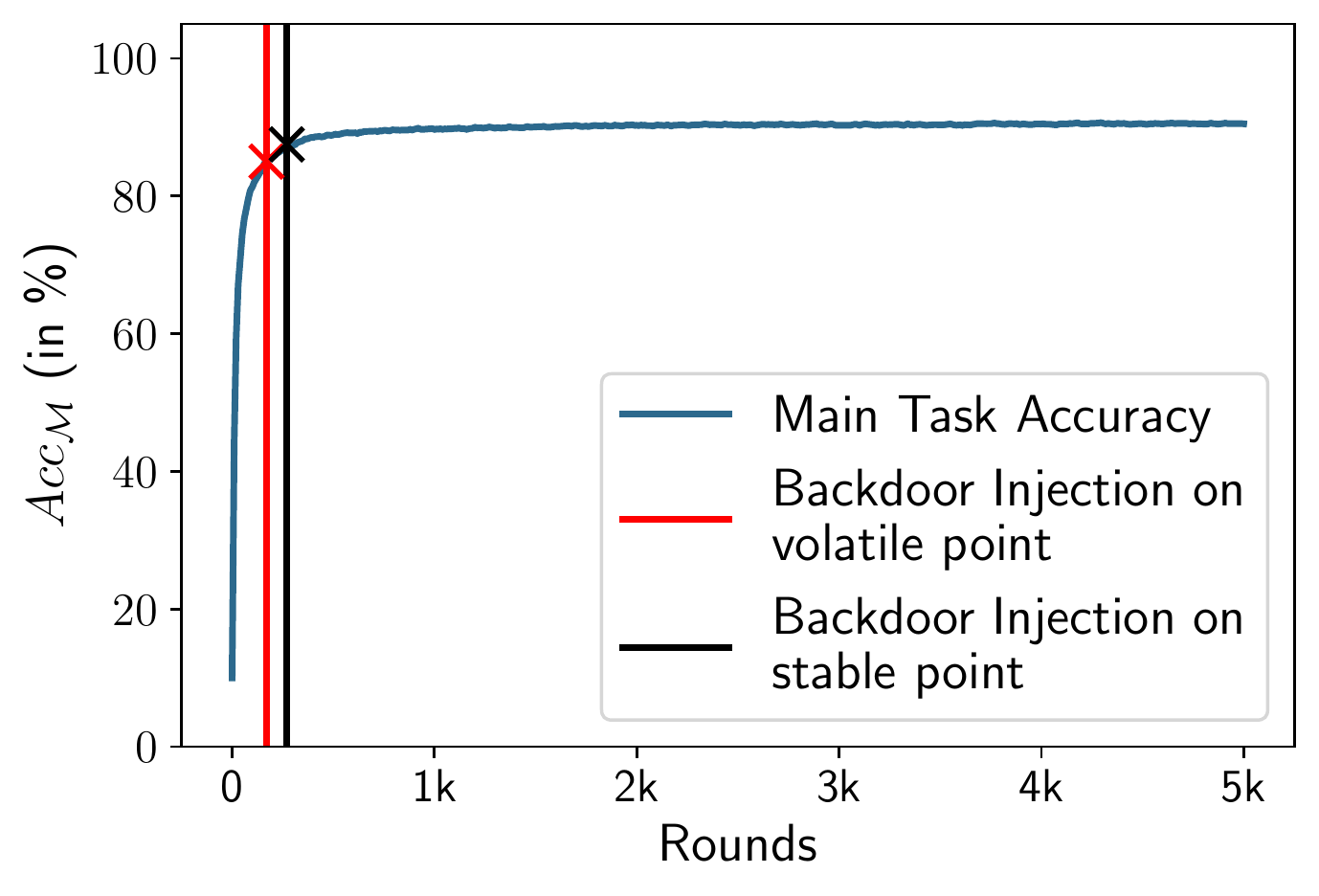}}
    \subfloat[\label{fig:fedavg_persistence}]{\includegraphics[width=0.58\linewidth]{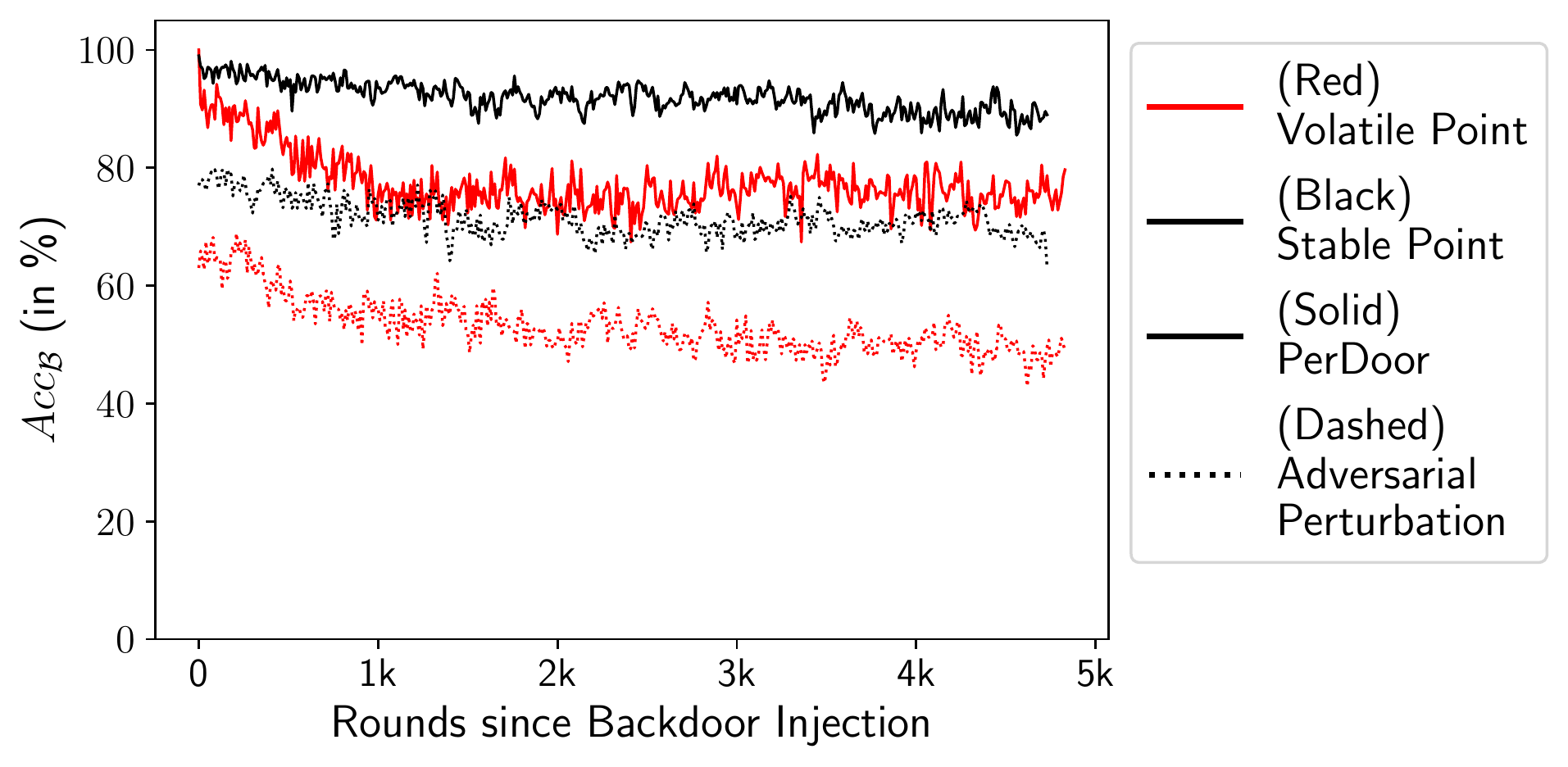}}
    \caption{(a) Global main task accuracy $Acc_{\mathcal{M}}$ in successive FL rounds and two instances of separate backdoor injection at volatile point and stable point. (b) Backdoor accuracy $Acc_{\mathcal{B}}$ in successive FL rounds after backdoor injection considering multiple scenarios.}
\end{figure}

\noindent \textbf{[RQ2] Persistence of \perdoor:}
We consider both the stable point and the volatile point, as discussed previously, during FL training for evaluating the persistence of injected backdoors. Figure~\ref{fig:fedavg_acc} shows $Acc_{\mathcal{M}}$ for $5000$ successive rounds where two separate instances of backdoors are injected at the volatile point and stable point. We can observe that the backdoor injection does not influence $Acc_{\mathcal{M}}$ (main task accuracy stays almost equivalent to when backdoors are not injected). We consider two scenarios to evaluate the persistence and effectiveness of \perdoor: \textbf{(1)} adversary injects backdoors only using adversarial examples but not restricting the effects on backdoor parameters, i.e., adversary does not employ Equation~\ref{eq:perdoor} for the local model modifications, and \textbf{(2)} adversary injects backdoors using \perdoor{}. Figure~\ref{fig:fedavg_persistence} shows $Acc_{\mathcal{B}}$ over successive FL rounds after the backdoor injection for both scenarios. Also, the adversary does not participate in further FL rounds once the backdoor is injected. We can observe that backdoors show higher persistence if they are injected at the stable point compared to the volatile point. We can also observe that only adversarial examples show reasonable persistence over successive FL rounds because of their high \textit{transferability}\footnote{Transferability represents the ability of an attacker against a deep learning model to be effective against a different, potentially unknown model~\cite{DBLP:conf/uss/DemontisMPJBONR19}.} property. However, integrated with \perdoor, backdoors display more prominent and persistent backdoor accuracy.

\begin{figure}[!t]
    \centering
    \subfloat[\label{fig:adv_example}]{\includegraphics[width=0.5\linewidth]{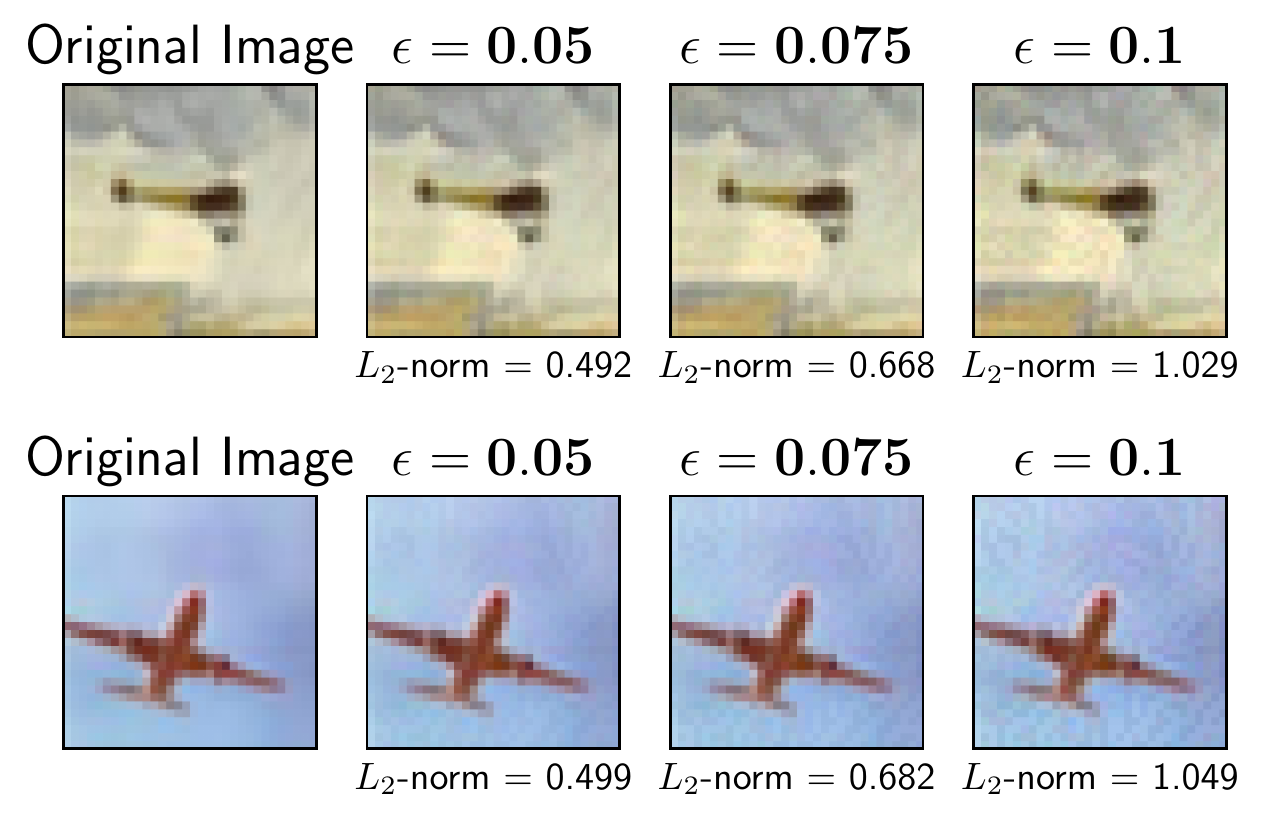}}
    \subfloat[\label{fig:adv_example_persistence}]{\includegraphics[width=0.5\linewidth]{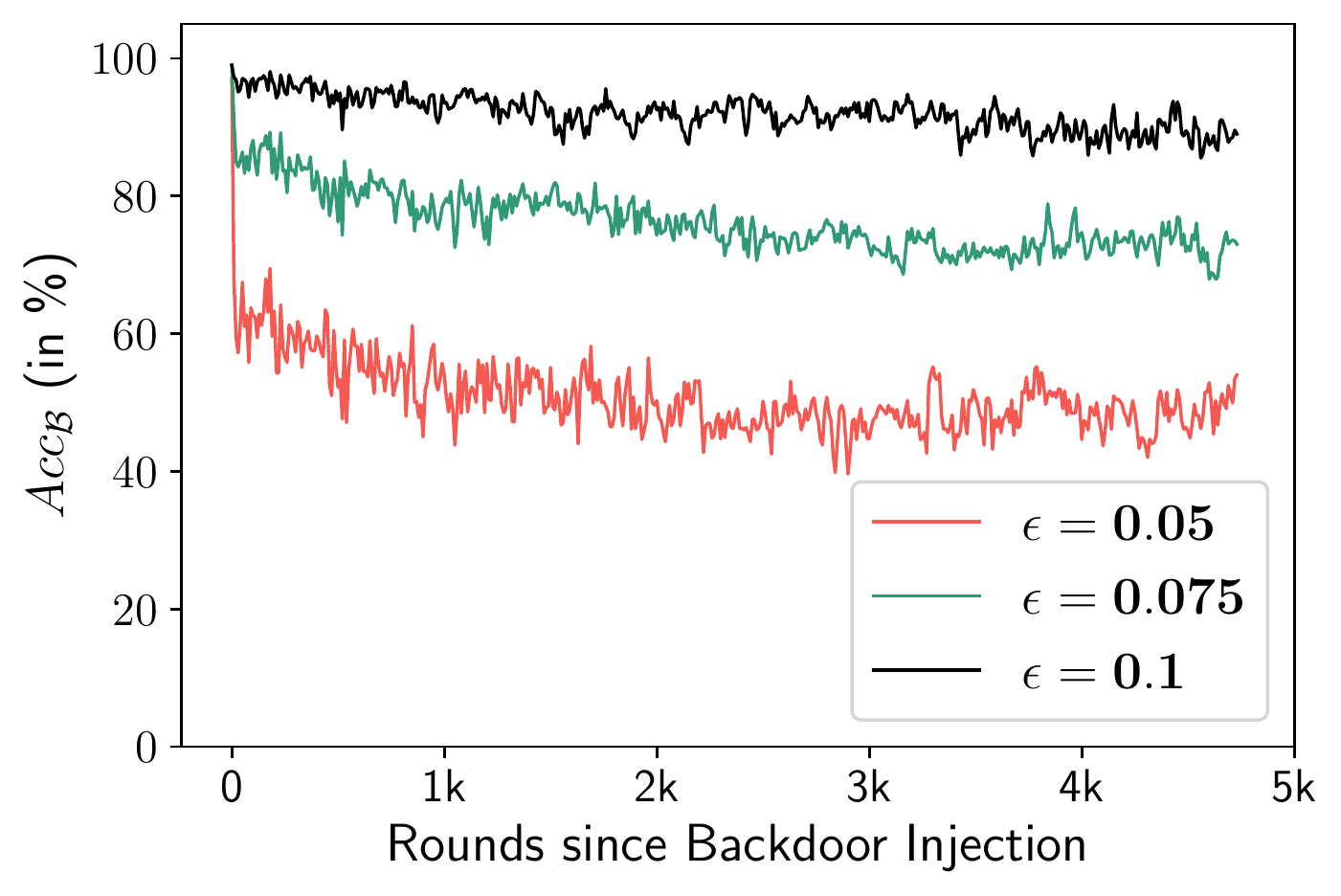}}
    \caption{(a) Backdoor images generated using adversarial perturbation with three different attack strength $\epsilon$ along with $L_2$-norm of the added perturbation. (b) Backdoor accuracy $Acc_{\mathcal{B}}$ in successive FL rounds after backdoor injection for different attack strengths.}
\end{figure}

We consider three different attack strengths $\epsilon$ for generating adversarial perturbations to evaluate the effectiveness of \perdoor{} against  different values of $\epsilon$. Figure~\ref{fig:adv_example} shows two benign images and corresponding adversarial images for attack strength $\epsilon = 0.05$, $\epsilon = 0.075$, and $\epsilon = 0.1$.
The figure also shows the $L_2$-norms of the noise added to benign images to generate adversarial counterparts, which act as triggers in the backdoor attack. As the attack strength increases, we can observe that images get more perturbed by strong noise patterns. Figure~\ref{fig:adv_example_persistence} shows $Acc_{\mathcal{B}}$ over successive FL rounds after the backdoor injection for different values of attack strength $\epsilon$, where the backdoor is injected at the stable point. We can observe that the backdoor persists for all these epsilon values; however, the backdoor accuracy is lower for weaker attacks.

To compare the persistence of \perdoor{} against other traditional backdoor injection techniques, we follow the methodology discussed in~\cite{DBLP:conf/aistats/BagdasaryanVHES20}. We consider a pre-defined set of backdoor images with
\begin{wrapfigure}{r}{0.5\linewidth}
  \begin{center}
    \includegraphics[width=\linewidth]{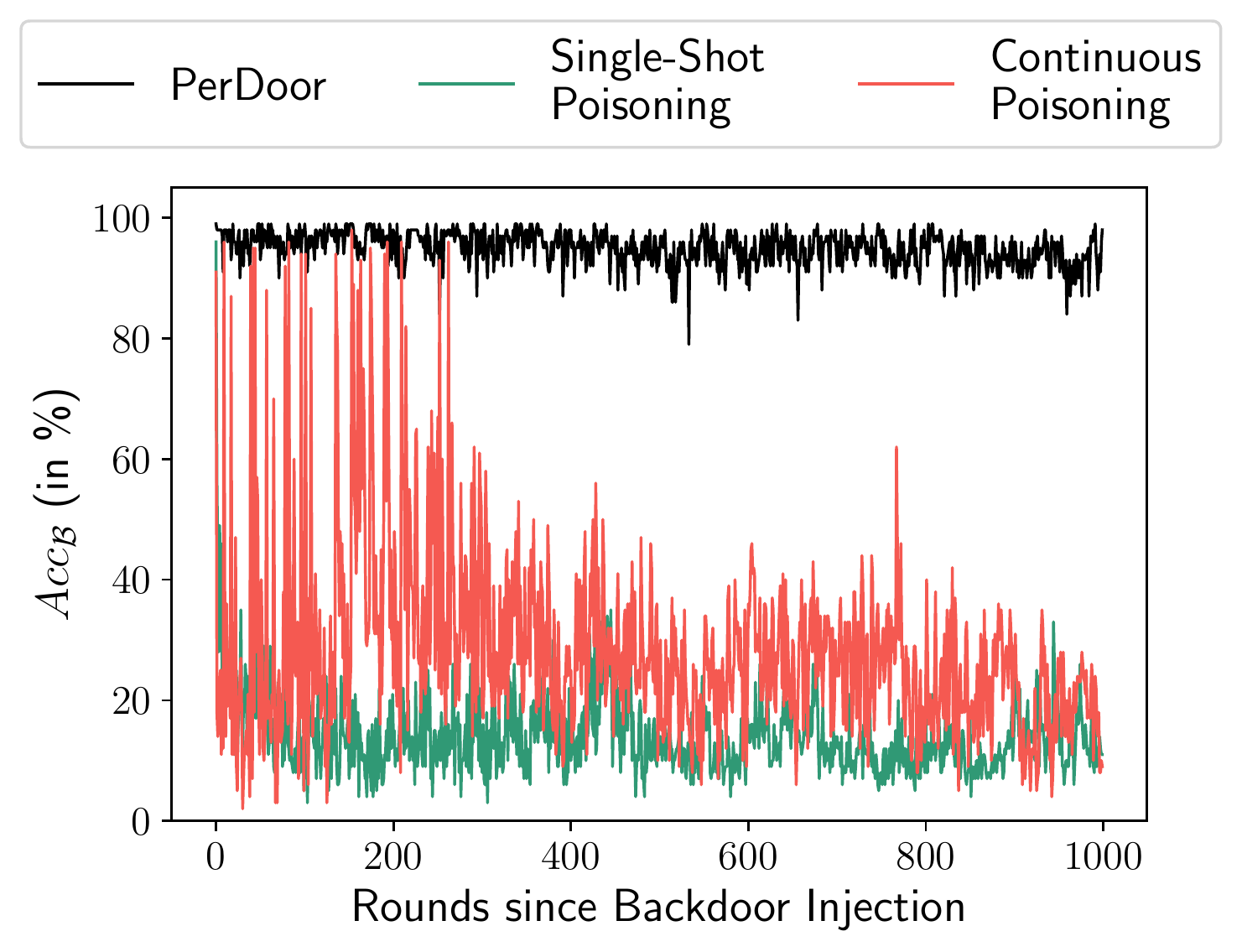}
  \end{center}
  \caption{Backdoor accuracy in successive FL rounds after backdoor injection following different methods.\label{fig:static_dynamic_persistence}}
\end{wrapfigure}
a fixed `cross' trigger pattern at the top left corner of the inputs. When selected, the adversary updates the broadcasted global model using backdoor images for 6 local epochs with an initial learning rate of $0.05$ that decreases by a factor of $10$ after every $2$ epochs and uploads the poisoned model for global aggregation. All these parameter values are chosen following the method discussed in~\cite{DBLP:conf/aistats/BagdasaryanVHES20}. We consider two scenarios for the evaluation: (1)~\textit{single-shot poisoning} - adversary injects backdoor at stable point, same as \perdoor{} and (2) \textit{continuous poisoning} - adversary injects backdoor continuously whenever it is selected during FL training from the beginning and stops uploading the poisoned updates after the stable point. Figure~\ref{fig:static_dynamic_persistence} shows $Acc_{\mathcal{B}}$ over successive FL rounds after the backdoor injection for both scenarios along with \perdoor. We can observe that single-shot poisoning obtains very high $Acc_{\mathcal{B}}$ immediately after that round. However, that decreases significantly as FL training continues. The continuous poisoning maintains high $Acc_{\mathcal{B}}$ at initial rounds when selected. However, the backdoor impact diminishes whenever the adversary stops uploading poisoned updates. On contrary, \perdoor{} maintains high $Acc_{\mathcal{B}}$ throughout the training rounds, demonstrating its efficacy in preserving $Acc_{\mathcal{B}}$. \perdoor, on average, achieves $10.5\times$ more $Acc_{\mathcal{B}}$ than single-shot poisoning over 5000 rounds and $6.4\times$ considering continuous poisoning.

\begin{figure}[!b]
    \centering
    \subfloat[\label{fig:acc_countermeasure}]{\includegraphics[width=0.5\linewidth]{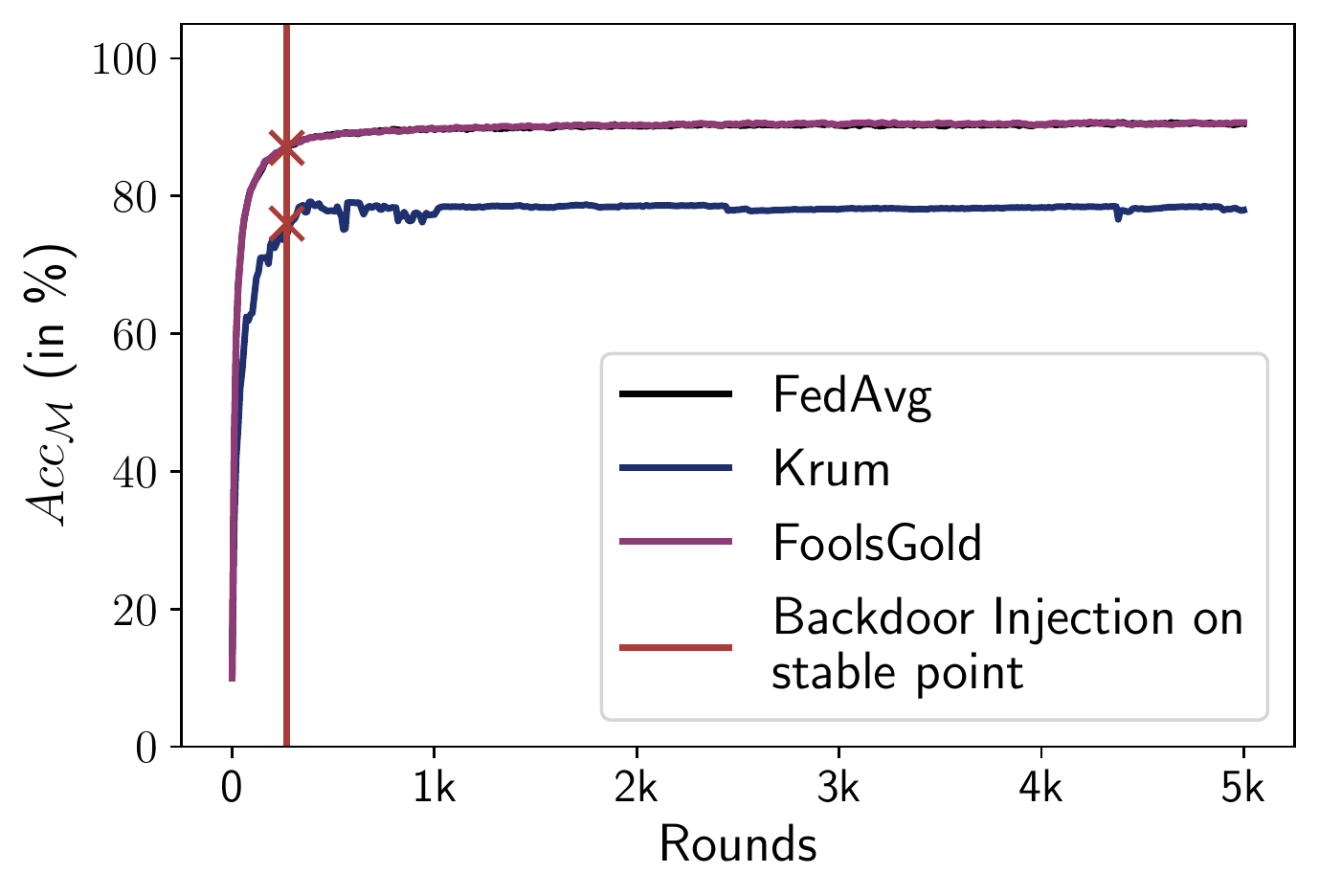}}
    \subfloat[\label{fig:persistence_countermeasure}]{\includegraphics[width=0.5\linewidth]{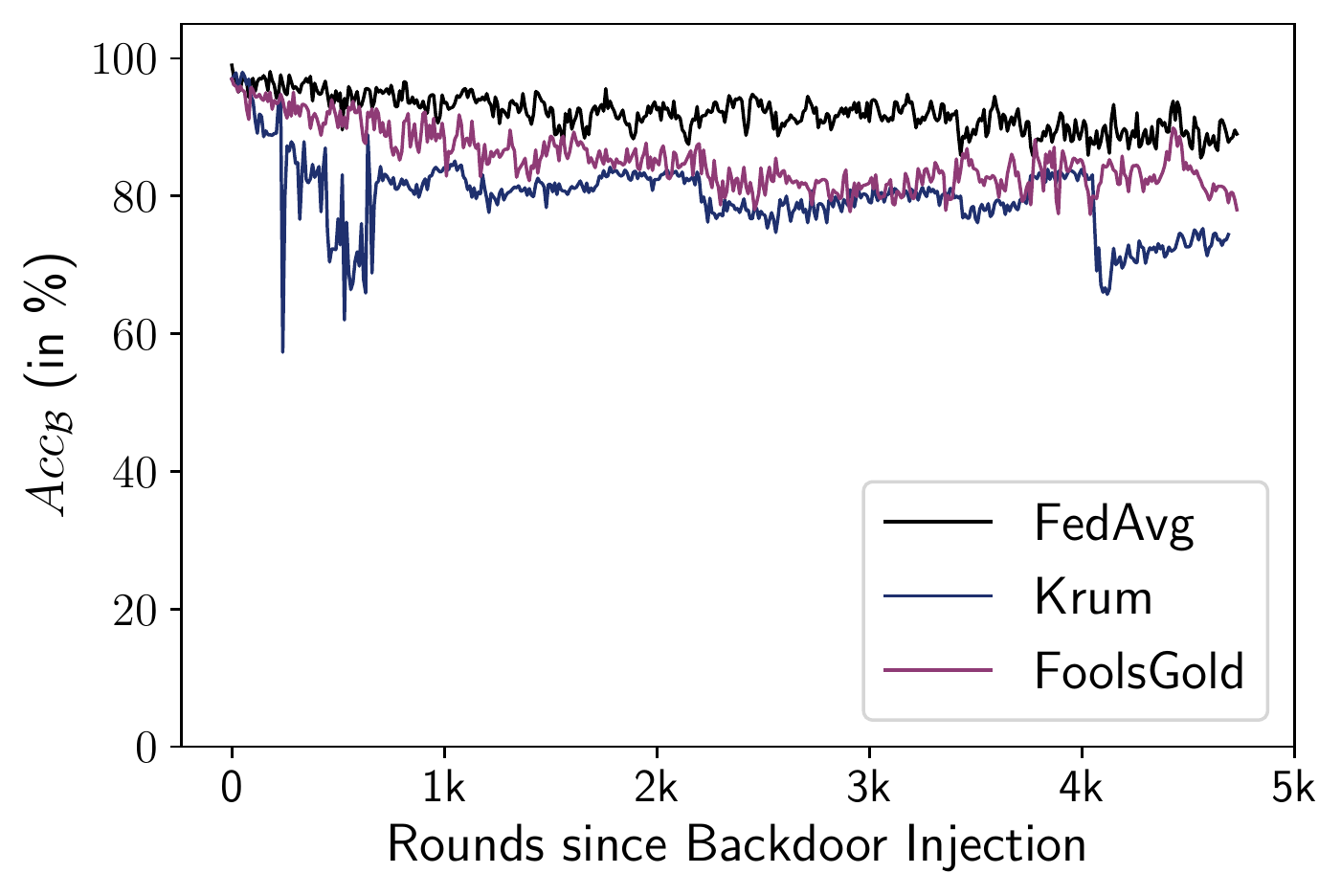}}
    \caption{(a) Global main task accuracy $Acc_{\mathcal{M}}$ in successive FL rounds and backdoor injection at stable point. (b) Backdoor accuracy $Acc_{\mathcal{B}}$ in successive FL rounds after backdoor injection considering different aggregations.}
\end{figure}

\noindent \textbf{[RQ3] Performance of \perdoor{} against state-of-the-art Defenses:}
We evaluate the effectiveness of \perdoor{} against two existing state-of-the-art defenses, Krum and FoolsGold. Figure~\ref{fig:acc_countermeasure} shows $Acc_{\mathcal{M}}$ for $5000$ successive rounds considering both these defenses and unprotected FedAvg. We consider that the adversary injects backdoors at the stable point. We can observe that while FoolsGold performs almost identical to unprotected FedAvg in terms of $Acc_{\mathcal{M}}$, the main task accuracy of Krum drops below 80\% (i.e., almost a drop of 10\%). It may be noted that the accuracy drop is not because of backdoor injection but due to the internal anomaly detection mechanism of Krum consistently producing global models representing data from the majority of clients. The result also aligns with the observations discussed in~\cite{DBLP:conf/aistats/BagdasaryanVHES20,DBLP:conf/raid/FungYB20}. Figure~\ref{fig:persistence_countermeasure} shows $Acc_{\mathcal{B}}$ over successive FL rounds after the backdoor injection considering all three aggregation methods. We can observe that \perdoor{} can efficiently inject impactful backdoors into the global model even in the presence of defense methods.

\begin{figure}[!t]
    \centering
    \subfloat[\label{fig:trigger_analysis}]{\includegraphics[width=0.7\linewidth]{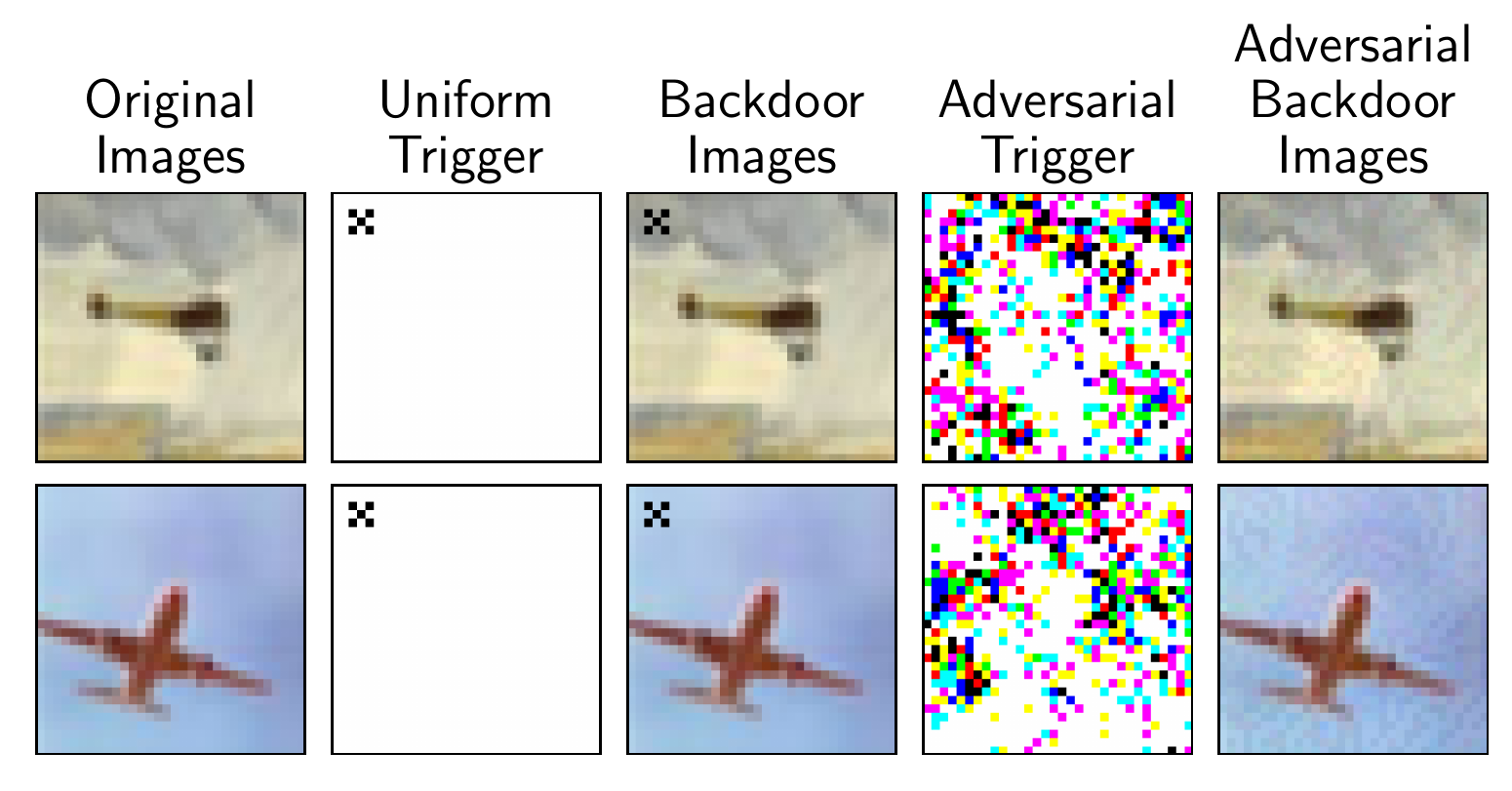}}
    \subfloat[\label{fig:trigger_ssd}]{\includegraphics[width=0.3\linewidth, height=4cm]{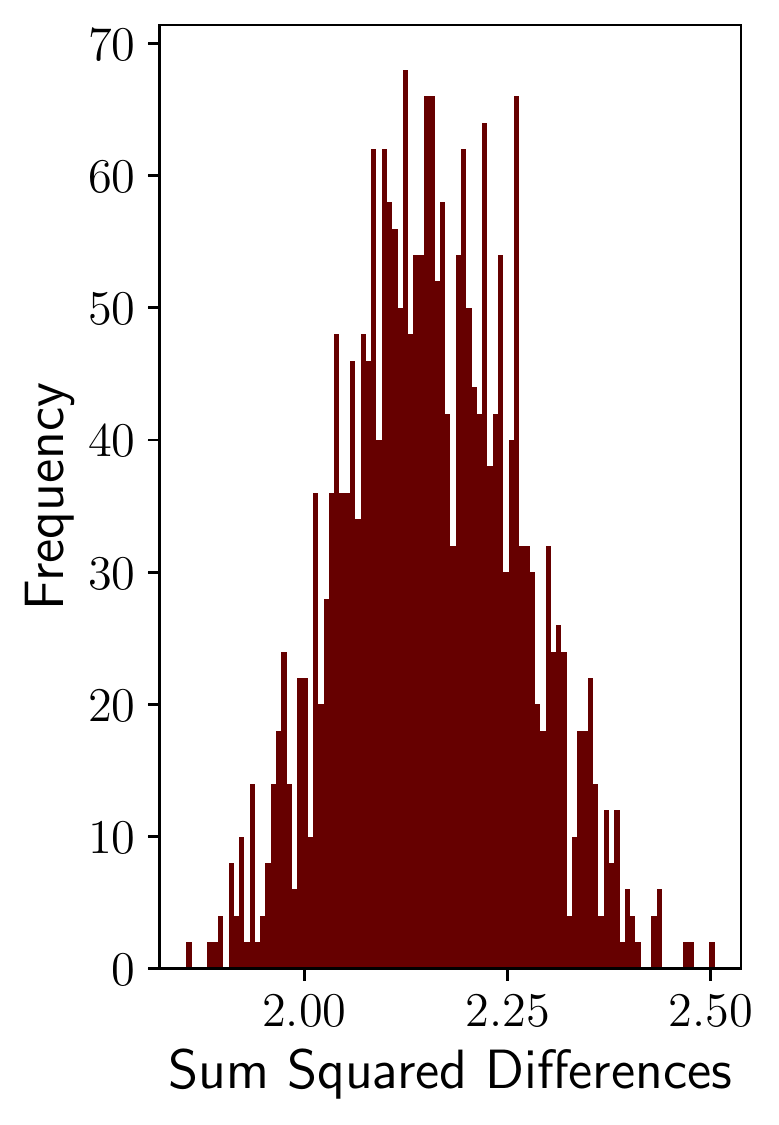}}
    \caption{(a) Backdoor images generated using uniform and adversarial trigger patterns. (b) Distribution of sum squared differences between each pair of $50$ adversarial triggers.}
\end{figure}

\noindent \textbf{[RQ4] Analysis of Trigger Patterns:}
We analyze the backdoor triggers generated using adversarial perturbation by \perdoor{} against the uniform trigger used in existing backdoor techniques~\cite{DBLP:conf/aistats/BagdasaryanVHES20,DBLP:conf/aaai/OzdayiKG21,DBLP:conf/iclr/XieHCL20}. Figure~\ref{fig:trigger_analysis} shows two instances of clean images belonging to the label `airplane', uniform trigger images with a fixed `cross' trigger pattern at the top left corner, and adversarial trigger images generated by \perdoor{}. The target label for both uniform and adversarial triggers is `frog'. Figure~\ref{fig:trigger_ssd} shows the distribution of pixel-wise sum squared differences between each pair of $50$ adversarial triggers. We can observe that the adversarial triggers generate a non-uniform distribution of patterns compared to one fixed value for the uniform counterpart considering the same source and target labels. The non-uniform patterns of adversarial triggers make it difficult for the central server to patch global models based on the knowledge of trigger patterns.

\section{Discussion}\label{sec:discussion}
\noindent \textbf{Limitations and Future Works:}
In this work, we consider: \textbf{(1)} an FL framework where the participants collectively train the global model using training data that is independent and identically distributed (IID) among the participants and \textbf{(2)} backdoor triggers that modify pixels of the input image. We keep the evaluation of \perdoor{} under a non-IID setting and more challenging-to-construct semantic backdoors (backdoors that do not tamper with input pixels) for future research exploration. We evaluated \perdoor{} in an image classification scenario using the CIFAR-10 dataset and VGG-11 architecture. However, we will also explore the same evaluation on different datasets, architectures, and application scenarios in our future research direction.

\noindent \textbf{Societal Impacts:} FL is specifically used in healthcare \cite{nature_fl} and its results may directly have financial impact (insurance) or societal stigma related to it. Persistent backdoor attacks in FL may enable an adversary exploit these consequences over a long period of time.

\section{Conclusion}
In this paper, we propose PerDoor, a persistent-by-construction backdoor injection method in FL targeting parameters of the global model that deviate less in successive FL rounds and contribute the least to the main task accuracy for injecting backdoors. PerDoor uses adversarial attacks to generate backdoor triggers from the global model, which not only helps to achieve good backdoor accuracy but also aids in generating non-uniform backdoor triggers for different inputs of the same class targeting the same label, compared to the uniform fixed triggers in existing backdoor methods in FL that are prone to be easily mitigated. Experimental evaluation considering an image classification scenario demonstrates that \perdoor{} can achieve, on average, $10.5\times$ persistence over successive iterations of FL compared to traditional backdoor injection methods when the backdoor is injected at the same round. \perdoor{} can also achieve high persistence even in the presence of state-of-the-art defenses.

\section*{Resources}
The implementation will be available at: \url{https://github.com/momalab/PerDoor} after the paper is published.
\bibliographystyle{plain}
\bibliography{main}

\appendix

\section{Brief Overview of Adversarial Example Generation}\label{sec:adv_attack}
Let us consider a benign data point $x \in \mathbb{R}^d$, classified into class $\mathcal{C}_i$ by a classifier $\mathcal{M}$. A targeted adversarial attack tries to add visually imperceptible perturbation $\eta \in \mathbb{R}^d$ to $x$ and creates a new data point $x_{adv} = x + \eta$ such that $\mathcal{M}$ misclassifies $x_{adv}$ into a target class $\mathcal{C}_j$. The imperceptibility is enforced by restricting the $l_{p}$-norm of the perturbation $\eta$ to be below a threshold $\epsilon$, i.e., $\|\eta\|_{p} \leq \epsilon$~\cite{DBLP:journals/corr/GoodfellowSS14, DBLP:conf/iclr/MadryMSTV18}. We term $\epsilon$ as the \textit{attack strength}.

An adversary can craft adversarial examples by considering the loss function of a model, assuming that the adversary has full access to the model parameters. Let $\mathcal{J}(\theta, x, y)$ denote the loss function of the model, where $\theta$, $x$, and $y$ represent the model parameters, benign input, and corresponding label, respectively. The goal of the adversary is to generate adversarial example $x_{adv}$ such that the loss of the model is minimized  for the target class $\hat{y}$ while ensuring that the magnitude of the perturbation is upper bounded by $\epsilon$. Hence, $\mathcal{J}(\theta, x_{adv}, \hat{y}) < \mathcal{J}(\theta, x, y)$ adhering to the constraint $\|x_{adv} - x\|_{p} \leq \epsilon$. Several methods have been proposed in the literature to solve this constrained optimization problem. We discuss the method used in the evaluation of our proposed approach.

\textbf{Basic Iterative Method (BIM):} This attack is proposed in~\cite{DBLP:conf/iclr/KurakinGB17a}. The adversarial example is crafted iteratively using the following equation
\begin{equation*}
    x^{(k+1)} = x^{(k)} - clip_{\epsilon}(\alpha \cdot sign(\nabla_x J(\theta, x^{(k)}, \hat{y})))
\end{equation*}
where $x^{(0)}$ is the benign example. If $r$ is the total number of attack iteration, $x_{adv} = x^{(r)}$. The parameter $\alpha$ is a small step size usually chosen as $\epsilon/r$. The function $clip_{\epsilon}(\cdot)$ is used to generate adversarial examples within $\epsilon$-ball of the original image $x^{(0)}$.

\end{document}